\documentclass[9pt,twocolumn,twoside]{osajnl}
\usepackage{yhmath}
\usepackage{graphicx}
\usepackage{subfigure}
\usepackage{upgreek}
\usepackage{inputenc}
\usepackage[format=plain,justification=justified]{caption}
\journal{ol} % Choose journal (ao, aop, josaa, josab, ol)
\setboolean{shortarticle}{true} 
\title{Fraunhofer diffraction at the two-dimensional quadratically distorted (QD) grating}

\author[1,2,$\dagger$]{Yuewei Liu}
\author[2,3,$\dagger$,$\ddagger$,*]{Yan Feng}

\affil[1]{School of mathematics and statistics, Lanzhou University, Gansu 730000, P. R. China}
\affil[2]{Scottish Universities Physics Alliance (SUPA), Institute of Biological Chemistry, Biophysics and Bioengineering, School of Engineering and Physical Sciences, Heriot-Watt University (HWU), Edinburgh EH14 4AS, UK}
\affil[3]{Beijing National Laboratory for Molecular Scienses, Key Laboratory of Molecular Nanostructure and Nanotechnology, Institute of Chemistry, Chinese Academy of Sciences (ICCAS), Beijing 100190, P. R.China}
\affil[$\dagger$]{These authors contributed equally to this work.}
\affil[$\ddagger$]{Present address: Laboratory BioEmergences (USR3695), CNRS, University Paris-Saclay, Gif-sur-Yvette 91198, France}
\affil[*]{Corresponding author: yanfeng4DMCMI@gmail.com.}

\ociscodes{(050.1970) Diffractive Optics; (050.1950) Diffraction Gratings; (110.4190) Multiple Imaging; (180.6900) Three-Dimensional Microscopy.}

\begin{abstract}
A two-dimensional (2D) mathematical model of quadratically distorted (QD) grating is established with the principles of Fraunhofer diffraction and Fourier optics. A discrete sampling method is applied for finding a numerical solution of the diffraction pattern of QD grating. An optimized working phase term, which determines the balanced energies and high efficiency of multi-plane images, can be obtained by the bisection algorithm. This 2D mathematical model allows the precise design of QD grating and improves the optical performance of simultaneous multi-plane imaging system.
\end{abstract}

\setboolean{displaycopyright}{true}

\begin{document}
\maketitle
\begin{figure}[htbp]
	\centering
	\fbox{\includegraphics[width=0.8\linewidth]{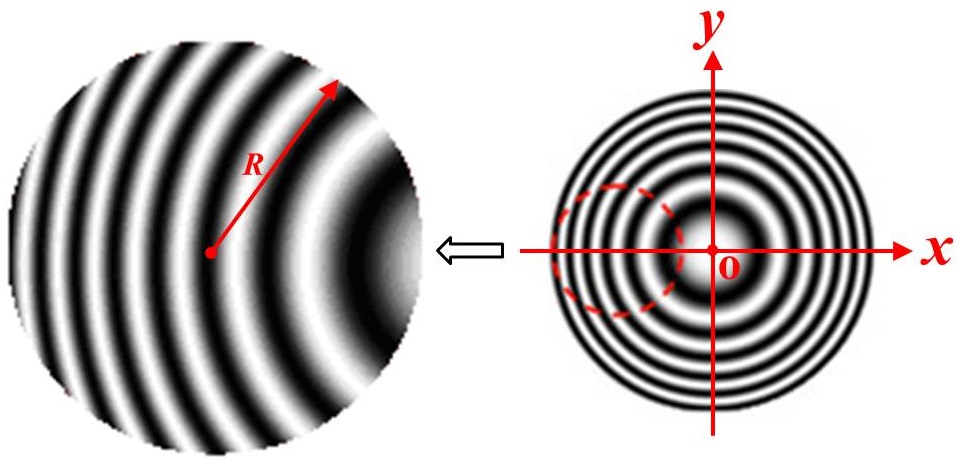}}
	\caption{ A demonstration of the structure of QD grating.}
	\label{Fig:QD}
\end{figure}

Using principles of adaptive optics, a novel simultaneous multi-plane imaging system based on quadratically distorted (QD) grating was developed by Greenaway and Blanchard in the 1990s \cite{Blanchard1999}, which was originally designed for photonic-crystal-fibre strain sensors and used in astronomy \cite{Greenaway2010}. QD grating, which is also known as an off-axis Fresnel Zone Plate (FZP) \cite{Born2001}, is formed by slits of a series of concentric circles with varying radii as demonstrated in Figure \ref{Fig:QD}. It behaves like a multi-focus ``lens'' but utilizes principle of diffraction instead of refraction, and provides an order-dependent focussing power to generate several images. This simple, on axis and scanless imaging system can simultaneously capture multiple, in-focus specimen planes on a single image plane, and z-plane separations of multi-focal images can be varied from arbitrarily small to many microns. As an optical attachment, it is fully compatible with a commercial microscope and standard camera system.  

A different implementation of similar principles, the so-called aberration-corrected multifocus microscopy (MFM), was developed by late Gustafsson and Abrahamsson \cite{Abrahamsson2013Fast}. It is capable of producing an instant focal stack of nine 2D images in multiple colours, and can be extended to image up to 25 focal planes under some circumstance \cite{Abrahamsson2015MultiFocus}. However, the multifocus grating (MFG), which is utilized to provide a variety of focal lengths (like QD grating), is designed by computer programming in a ``black box'' instead of theoretical analysis \cite{Abrahamsson2016Multifocus}. Due to the technical ceilings, this customized optical system may only be appropriate for limited applications.

To develop a versatile and easy-to-use simultaneous multi-plane imaging system, in particular in terms of the varied z-separations between object planes and considerable large field of view (FOV), an analytically designed QD grating is essential. However, due to the high level of sophistication in 2D mathematical modeling of QD grating, in the past a rough one-dimensional (1D) model was built for the design and optimization of QD grating, in which the characteristic quadratic curvature and chirped-period were simplified as 1D equidistant slits \cite{Yan2013}. In this paper, we will establish an elaborate 2D mathematical model of QD grating, calculate both analytic and numerical solutions of the Fraunhofer diffraction pattern, and finally verify this model.

It has been proved that the scalar diffraction theory can be utilized for evaluating the image formed with a source of natural light by an optical system of moderate numerical aperture, thus an approximate description in terms of a single complex scalar wave function is adequate to describe most problems encountered in optics.  The further approximation, which is referred to as Fraunhofer approximation, has greatly simplified the calculations of diffraction patterns under certain conditions. The evaluation of Fraunhofer diffraction can be written as a Fourier integral  \cite{Born2001} 
\begin{equation}\label{Eq:Fraun}
	U(p,q)=\iint G(x,y)e^{-\frac{2\pi i}{\lambda}(px+qy)}\mathrm{d}x \mathrm{d}y,
\end{equation}
where $\lambda$ is the incident wavelength. The integral extends over the whole $x$-$y$ plane, and $G(x, y)$ the pupil function is given by
\begin{equation}\label{Eq:QDFun}
	G(x,y)=
	\begin{cases}
		\text{constant}, &\text{ at points in the opening},\\
		0, &\text{ at points outside the opening.}
	\end{cases}
\end{equation}

\begin{figure}[htbp]
		\centering
		\includegraphics[width=0.5\textwidth]{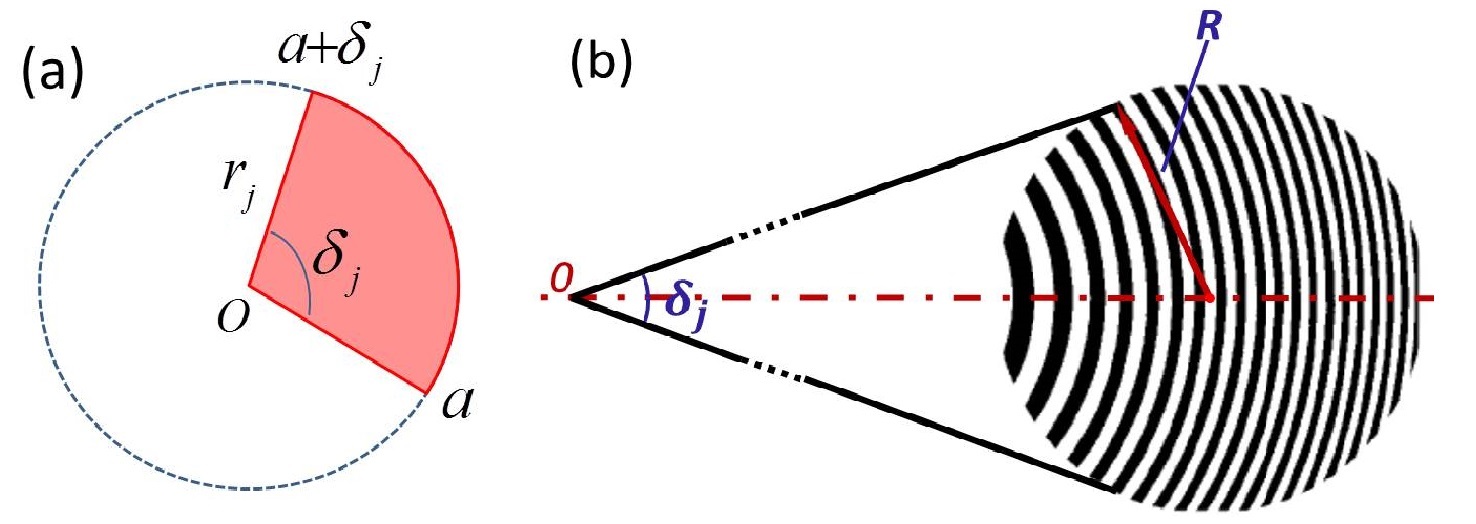}
	%	\captionsetup{format=plain,justification=justified}
	\caption{(a) Parameters of the $j$th circular sector: $a$ is the start point, $\delta_j$ the central angle and $r_j$ the radius. (b) Schematic of a QD grating pattern with radius $R$ and the $j$th slit.}
	\label{Fig:Fig2}
\end{figure}

\begin{figure}[htbp]
	\centering
	\fbox{\includegraphics[width=0.9\linewidth]{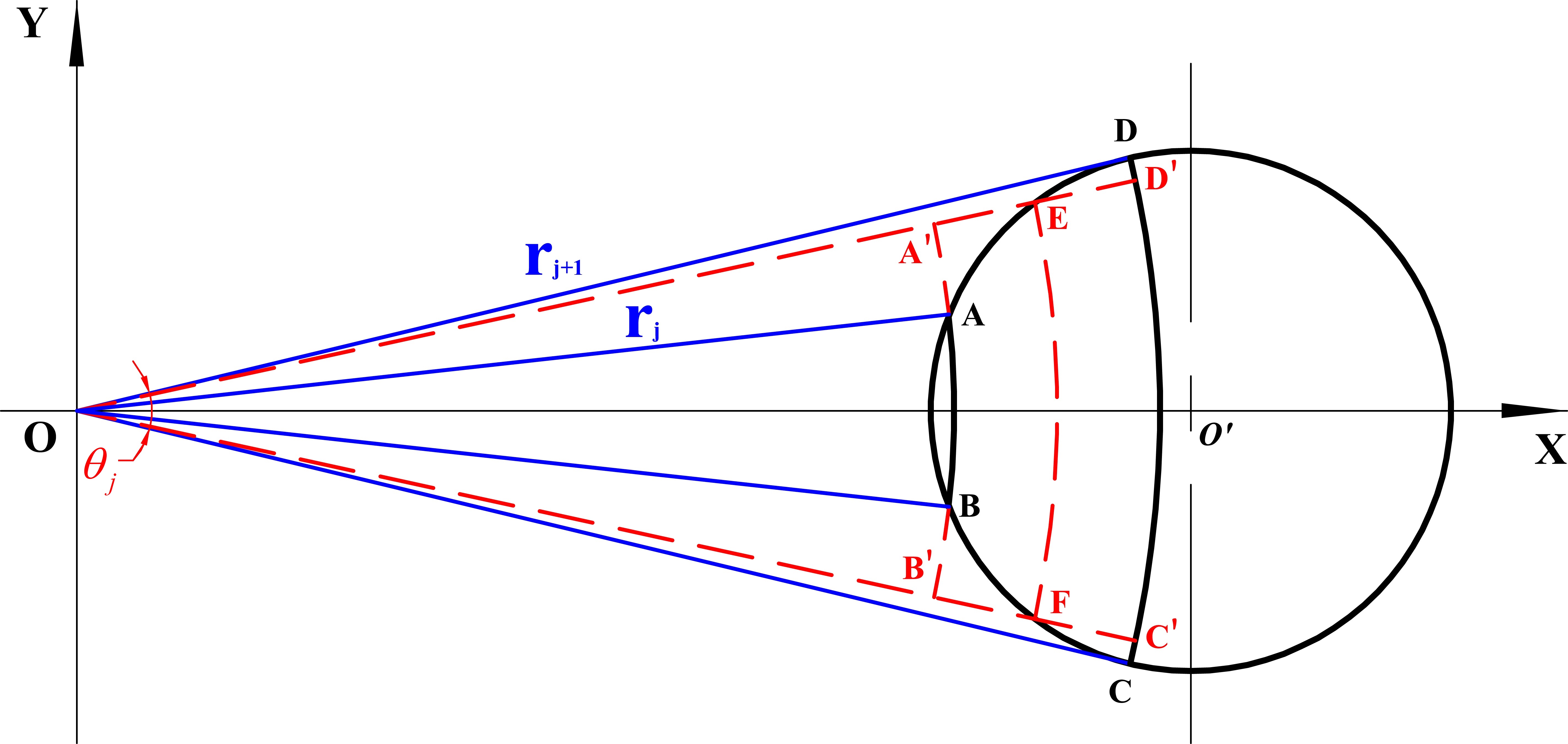}}
	\caption{Illustration of an approximation of a single slit in 2D QD grating.}
	\label{Fig:SglSlit}
\end{figure}

Hence the complex field distribution across the Fraunhofer diffraction pattern can be simply obtained by the Fourier transform of the aperture function \cite{Goodman2005}. However, due to the complicated slit profile of QD grating, an alternative way is implemented. Based  on the linearity and superposition properties of Fourier transform, Fraunhofer diffraction pattern of a QD grating is presented by the linear superposition of the Fourier spectra of all the concentric slits, which are calculated by the subtraction of the Fourier spectra between adjacent sectors (see Figures \ref{Fig:Fig2} \& \ref{Fig:SglSlit}).  The effect of the marginal fractions, which is generated by the subtraction of adjacent sectors, can be neglected due to their limited contribution to the diffraction images.

Now attention has been focussed on the Fourier transform of a single sector as shown in Figure \ref{Fig:Fig2}(a). Consider a QD grating comprising of $N$ slits (thus $N+1$ sectors), the Fourier integral of the $j$th sector can be written as
\begin{equation}\label{Eq:ItgSct}
	F(p,q;r_j,\delta_j)=\iint G_j(x,y) e^{-\frac{2\pi i}{\lambda}(px+qy)}\mathrm{d}x \mathrm{d}y,
\end{equation}
where $G_j(x, y)$ corresponds to the $j$th sector in the QD grating with a radius of $r_j$ and central angle of $\delta_j$ starting at point $a,$ and is set to be 1 in the integral domain of the sector. The integer $j$ varies from negative to positive values and $j = 0$ represents the arc that passes through the grating centre.

To exploit the circular symmetry of $G_{j}(x, y),$ a transformation to polar coordinates in both the $(x, y)$ and the $(p, q)$ planes is made as follows:
\begin{equation}\label{Eq:PolarTrans}
	\begin{cases}
		r=\sqrt{x^2+y^2}, \ \theta=\arctan(y/x),\\
		\rho=\sqrt{p^2+q^2}, \ \phi=\arctan(q/p).
	\end{cases}
\end{equation}

Applying the coordinate transformations (\ref{Eq:PolarTrans}) to (\ref{Eq:ItgSct}), the Fourier integral of the $j$th sector becomes \cite{Goodman2005}.
\begin{eqnarray}\label{Eq:ItgSctPolar}
	F(\rho,\phi;r_j,\delta_j)&=&\frac{e^{ikz}e^{ik\rho^2/2z }}{i\lambda z}\int_0^{r_j}\int_a^{a+\delta_j} e^{i\frac{2\pi}{\lambda z}\rho r \sin (\theta-\phi-\frac{\pi}{2})}r\mathrm{d}r \mathrm{d}\theta \nonumber\\ 
	&=&\frac{e^{ikz}e^{ik\rho^2/2z }}{i\lambda z}L,
\end{eqnarray}
where $k=2\pi/\lambda$ is the number of waves, and $z$ the observation distance between aperture and image plane, which is approaching the far-field conditions of Fraunhofer diffraction. 

By the Jacobi-Anger expansion, 
\begin{equation}\label{Eq:JacbobiExp}
	e^{i\beta \sin \alpha}=\sum_{n=-\infty}^{+\infty}J_n(\beta) e^{in\alpha},
\end{equation}
where $J_n(\cdot)$ is the first kind Bessel function of order $n,$
the integral $L$ in  (\ref{Eq:ItgSctPolar})  becomes \vspace{-0.2cm}
\begin{equation}\label{Eq:K}
	L=\sum_{n=-\infty}^{+\infty}\int_0^{r_j}J_n\left(\frac{2\pi}{\lambda z}\rho r\right)r \mathrm{d}r\int_0^{\delta_j}e^{in(\theta+\psi)}\mathrm{d}\theta,
\end{equation}
where $\psi=(a-\phi-\frac{\pi}{2}).$
Since the first kind Bessel function follows that
\begin{equation}
\int_0^x \zeta J_0(\zeta) \mathrm{d}\zeta=xJ_1(x)
\end{equation}
and 
\begin{equation}
J_{-n}(x)=(-1)^nJ_n(x),
\end{equation}
(\ref{Eq:K}) can be written in the form
\begin{equation}\label{Eq:K0}
	L_0=\frac{\delta\lambda z r_j}{2\pi \rho}J_1\left(\frac{2\pi}{\lambda z}\rho r_j\right)
\end{equation}
when $n=0,$ and\vspace{-0.3cm}
\begin{equation}\label{Eq:Knew}
	L=L_0+\left(\frac{2\pi}{\lambda z}\rho\right)^{-2}\sum_{n=1}^{+\infty}L_n,
\end{equation}
where 
\begin{eqnarray}\label{Eq:Kn}
	L_n&=&\int_0^{\frac{2\pi\rho r_j}{\lambda z}}rJ_n(r)\mathrm{d} r\int_0^{\delta_j}\left[e^{in(\theta+\psi)}+(-1)^n e^{-in(\theta+\psi)}\right]\mathrm{d}\theta \nonumber \\
	&=& \Psi_n\Phi_n.
\end{eqnarray}

Further we have \cite{Online2015}
\begin{equation}\label{Eq:CB}
	\int_0^xt^{\mu}J_v(t)\mathrm{d} t= x^\mu C \sum_{k=0}^{+\infty}C_k(x),
\end{equation} 
where \vspace{-0.1cm}
\begin{equation}
	C=\frac{\Gamma\left(\frac{1}{2}v+\frac{1}{2}\mu+\frac{1}{2}\right)}{\Gamma\left(\frac{1}{2}v-\frac{1}{2}\mu+\frac{1}{2}\right)},
\end{equation}
and 
\begin{equation}
	C_k(x)=(v+2k+1)\frac{\Gamma\left(\frac{1}{2}v-\frac{1}{2}\mu+\frac{1}{2}+k\right)}{\Gamma\left(\frac{1}{2}v+\frac{1}{2}\mu+\frac{3}{2}+k\right)}J_{v+2k+1}(x),
\end{equation}
where $\Gamma(\cdot)$ denotes the Gamma function.

Let $\mu=1$ and $x=2\pi\rho r_j/\lambda z,$ and substitute these into (\ref{Eq:CB}),  $\Psi_n$ in (\ref{Eq:Kn}) can be given by\vspace{-0.1cm}
\begin{equation}\label{Eq:PsiEven}
	\Psi_{2m}\!=\!\frac{2\pi}{\lambda z}\rho r_j m\sum_{k=0}^{+\infty}\frac{2m+2k+1}{(m+k)(m+k+1)}J_{n+2k+1}\left(\frac{2\pi}{\lambda z}\rho r_j\right),
\end{equation}
\vspace{-0.1cm}for an even number $n=2m,$  and\vspace{-0.1cm}
\begin{equation}\label{Eq:|PsiOdd}
	\begin{split}
		\Psi_{2m-1}=&2(2m-1)\frac{2\pi}{\lambda z}\rho r_j m\cdot\\ 
		&\sum_{k=0}^{+\infty}\frac{m+k}{(m+k+\frac{1}{2})(m+k-\frac{1}{2})}J_{n+2k+1}\left(\frac{2\pi}{\lambda z}\rho r_j\right),
	\end{split}
\end{equation}
for an odd number $n=2m-1.$

The second integral $\Phi_n$ in (\ref{Eq:Kn}) can be expressed as
\begin{equation}\label{Eq:PhiEven}
	\Phi_{2m}=\frac{2(-1)^m}{m}\sin (m\delta_j) \cos (m\delta_j-2m\phi+2ma)
\end{equation}
for an even number $n=2m,$
and 
\begin{equation}\label{Eq:PhiOdd}\
	\begin{split}
		\Phi_{2m-1}=\frac{4i(-1)^m}{2m-1}\sin \left[\frac{(2m-1)\delta_j}{2}\right] \cos \left[\frac{(2m-1)(\delta_j+2a-2\phi)}{2}\right]
	\end{split}
\end{equation}
for an odd number $n=2m-1.$ 

Based on the expressions of  $(L_0,\Psi_{2m},\Psi_{2m-1},\Phi_{2m},\Phi_{2m-1})$ defined in (\ref{Eq:K0},\ref{Eq:PsiEven}-\ref{Eq:PhiOdd}) respectively, we finally have the Fourier integral of the $j$th sector
\begin{equation}\label{Eq:Fj}
	\begin{split}
		&F(\rho,\phi;r_j,\delta_j)=\frac{e^{ikz}e^{ik\rho^2/2z }}{i\lambda z}\cdot \\  
		& \left[L_0+\left(\frac{2\pi}{\lambda z}\rho\right)^{-2}\sum_{m=1}^{+\infty}(\Psi_{2m}\Phi_{2m}+\Psi_{2m-1}\Phi_{2m-1})\right].
	\end{split}
\end{equation}

To obtain the Fourier integral (thus the Fraunhofer diffraction pattern) of QD grating, a linear superposition of the Fourier spectra of all the concentric slits is applied according to the linearity property of Fourier transform
\begin{equation}\label{Eq:SumUj}
	U(p,q)=\sum_{j\in  J} \iint S_j(x,y)e^{-\frac{2\pi i}{\lambda}(px+qy)}\mathrm{d}x \mathrm{d}y=\sum_{j\in J}U_j,
\end{equation}
where  $S_j$ represents the $j$th slit and $J =\{\text{integer } j:\lceil{-N/2}\rceil \le j \le \lceil {N/2}\rceil \}.$ 

We now consider arbitrary two neighboring sectors $\wideparen{OBA}$ and $\wideparen{OCD}$ as shown in Figure \ref{Fig:SglSlit}. The coordinate origin $O$ is the centre of the adjacent arcs $\wideparen{BA}$ and $\wideparen{CD}.$ If a virtual arc $\wideparen{FE},$ whose radius is the mid-value of that of arcs $\wideparen{BA}$ and $\wideparen{CD},$ is defined, then the Fourier integral of the $j$th slit $\wideparen{ABCD}$ can be approximated by that of an alternative domain $\wideparen{A^{\prime} B^{\prime} C^{\prime} D^{\prime}}$ when $r_j$ is much larger than the period of QD grating. Furthermore, the Fourier integral of the virtual slit $\wideparen{A^{\prime} B^{\prime} C^{\prime} D^{\prime}}$ can be approximated  by the subtraction of Fourier integrals of the two virtual adjacent sectors $\wideparen{O B^{\prime} A^{\prime}}$ and $\wideparen{O  C^{\prime} D^{\prime}}.$ Therefore, the Fourier integral of the $j$th slit can be estimated by
\begin{equation}\label{Eq:Slitj}
U_j\approx \iint \left(G_{\wideparen{O  C^{\prime} D^{\prime}}}-G_{\wideparen{O  B^{\prime} A^{\prime}}}\right)e^{-\frac{2\pi i}{\lambda}(px+qy)}\mathrm{d}x \mathrm{d}y.
\end{equation}	

With (\ref{Eq:Fj}) and (\ref{Eq:Slitj}), the Fourier spectrum of the $j$th slit is consequently estimated by
\begin{equation}\label{Eq:UjRes}
	U_j\approx F(\rho,\phi;r_{j+1},\theta_j)-F(\rho,\phi;r_j,\theta_j), \ \text{for } j=1,2,\ldots,N,
\end{equation}
where $\theta_j$ is the central angle of the $j$th virtual sector $\wideparen{O F E}$ shown in Figure \ref{Fig:SglSlit}. On substituting from (\ref{Eq:UjRes}), the Fourier spectrum of QD grating (\ref{Eq:SumUj}) can be re-written as
\begin{equation}\label{Eq:U}
	\begin{split}
		U_{Q}(p,q)=&U_{Q}(\rho\cos \phi,\rho \sin \phi)\\
		\approx &\sum_{j=1}^{N}e^{i\omega\cdot bool }(F(\rho,\phi;r_{j+1},\theta_j)-F(\rho,\phi;r_j,\theta_j)),
	\end{split}
\end{equation}
where $\omega$ is the phase shift, and $bool$ alternates between 1 and 0 with respect to $j$ is odd- and even- number respectively.

To find a numerical solution of the Fourier spectrum of QD grating (\ref{Eq:U}), a discrete sampling method is applied, such that the complexity of high order Bessel functions and tremendous series can be reduced. The images which correspond to the diffraction domains of first three orders (simultaneous 3-plane imaging in this case) are estimated by the method, respectively. Then the sampling is performed with certain intervals in radius $(\rho)$ and argument $(\phi)$ dimensions, which highly depends on the parameters of QD grating. Finally, a cubic interpolation algorithm is applied to smooth the spectrum surface. As an example, we set $z=10/ \lambda$ when a QD grating shown in Table \ref{Tbl:ParaQD} is implemented.

%\section{Numerical solution of Fraunhofer diffraction pattern}
%\textcolor{red}{Beginng of rewriting}\\
According to the Parseval’s theorem \cite{Goodman2005}, the absolute square of Fourier spectrum presents the energy distribution of Fraunhofer diffraction pattern. So the intensity of images over the whole imaging domain can be expressed as
\begin{equation}
E(\omega)=\iint |U_{Q}(\rho\cos \phi,\rho\sin \phi)|^2 \mathrm {d}\rho \mathrm{d} \phi.
\end{equation}

We have proved that the phase shift generated by the different etch depth of grating determines the percentage of the incident light that is directed into each diffraction order \cite{Yan2013}. For our 3D imaging system based on QD grating, the working phase (thus the target etch depth) can lead to the desired intensity balance between multi-plane images in each diffraction order as well as the maximum total energy in those orders. Ideally, on the assumption that all the incident flux is focussed only on the first three diffraction orders, we set $E_0(\omega)=E_{\pm 1}(\omega)=1/3E(\omega),$ where $E_0(\omega)$ and $E_{\pm 1}(\omega)$ denote the image intensities of zeroth and first orders, respectively. Since the two images of first orders are identical, we only take one of them into account. Consequently, the working phase is the root of
\begin{equation}
C(\omega)=\frac{1}{3}E(\omega)-E_0(\omega).
\end{equation}

\begin{algorithm}
	\caption{The bisection algorithm}\label{Alg:bisec}
	\begin{algorithmic}[1]
		\While{$|\omega_1-\omega_2|>\epsilon$ and $C(\omega)\neq 0, \text{where } \omega=\frac{\omega_1+\omega_2}{2},$ } \vspace{-0.1cm}
		\begin{eqnarray*}
			\omega_1&=&\omega, \text{ if } C(\omega_1)C(\omega)>0, \\  \vspace{-0.2cm}
			\omega_2&=&\omega, \text{ otherwise,}
		\end{eqnarray*}  
		\EndWhile \vspace{-0.3cm}
		\State \textbf{return} $\omega.$
	\end{algorithmic}
\end{algorithm}
Bisection algorithm \cite{Burden1985}, which is a root-finding method for a continuous function $C(\omega),$ typically works with two initial guesses, $\omega_1$ and $\omega_2$, such that $C(\omega_1)$ and $C(\omega_2)$ have opposite signs and at least one root can be bracketed within a subinterval of $[\omega_1,\omega_2]$ according to the intermediate value theorem. As  Algorithm \ref{Alg:bisec} shown, the interval between $\omega_1$ and $\omega_2$ will become increasingly smaller, converging on the root of the function after a few iterations.  Here the tolerance $\epsilon>0$ can reach up to $10^{-6}.$

A QD grating with moderate parameters (as Table \ref{Tbl:ParaQD} shown) is selected and applied in both our 2D model and the 1-D and period-fixed grating model \cite{Yan2013}, such that the two values of working phase obtained by both models should be close to each other.
\begin{table}[h]
	\renewcommand\arraystretch{1.2}
	\setlength{\belowcaptionskip}{1pt}
	\centering
	\caption{\bf Designed parameters of the 2D QD grating}
	\begin{tabular}{cc}
		\hline
		Central Period ($d_0$) & $50\upmu \rm{m}$ \\
		Radius ($R$) & $10$mm \\
		$W_{20}$ & $50\lambda$\\
		Wavelength ($\lambda$) & $532$nm\\
		Number of Arcs &$801$ \\
		\hline
	\end{tabular}
	\label{Tbl:ParaQD}
\end{table}     
\\where $W_{20}$ is the standard coefficient of defocus and is equivalent to the extra path length introduced at the edge of the aperture. And the varying radii $r_j,$   for $ j\in J,$ can be obtained by \cite{Blanchard1999}
\begin{equation}
r_j=\left[ \frac{j\lambda R^2}{W_{20}}+\left(\frac{\lambda R^2}{2d_0 W_{20}}\right)^2 \right ]^{1/2}.
\end{equation}

\begin{figure}[htb]
	\subfigure{
		\hspace{0.5cm}
		\includegraphics[width=0.4\textwidth]{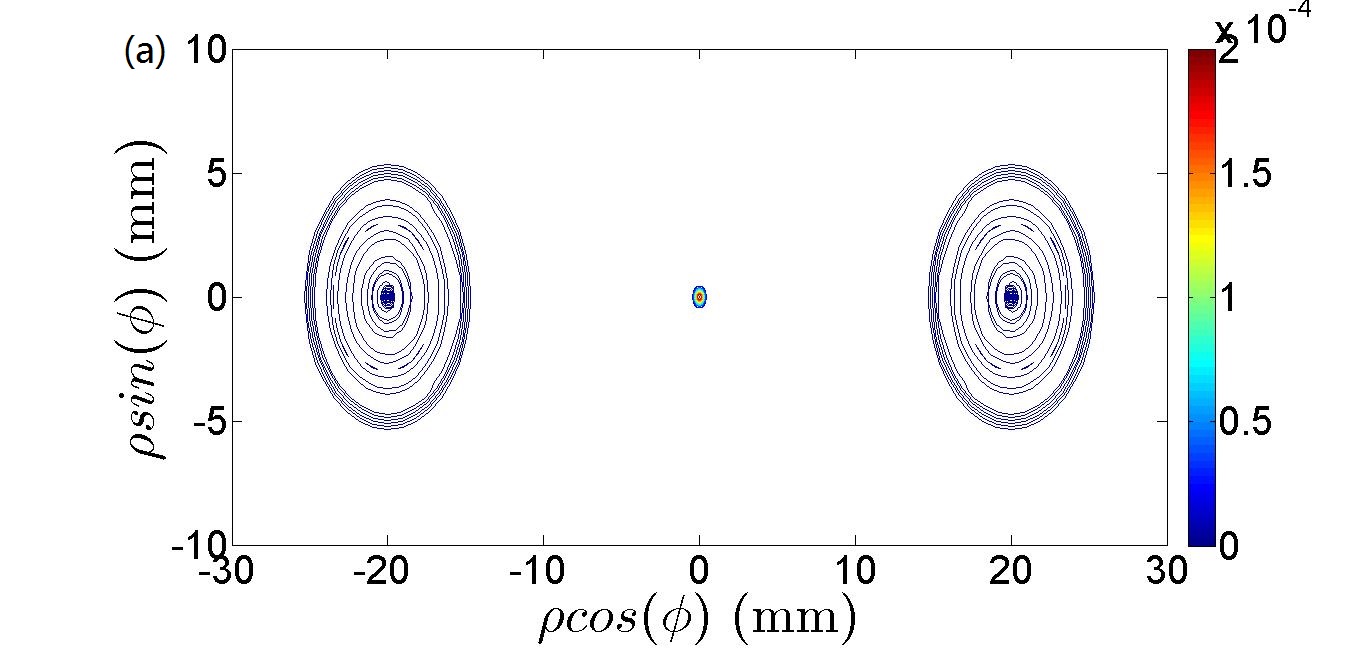} 
	}\\[-0.35cm]
	\subfigure{
		\hspace{0.5cm}
		\includegraphics[width=0.4\textwidth]{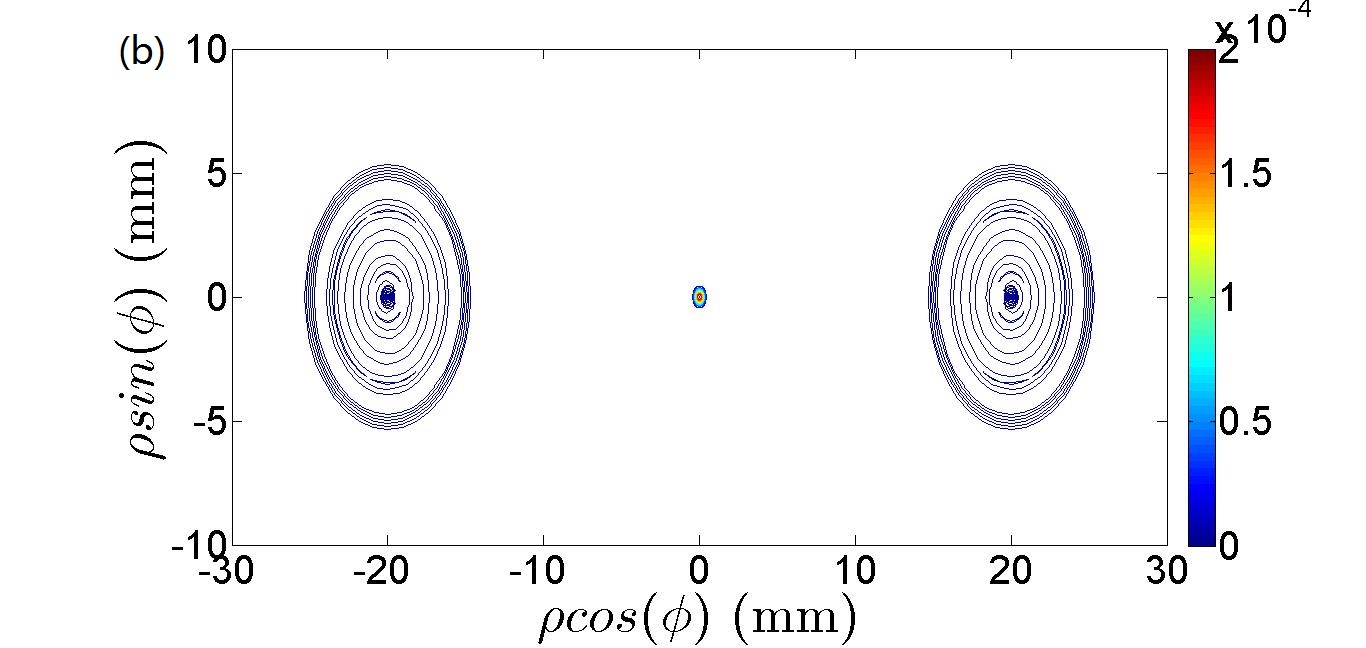}
	}
	\caption{Contour plot of normalized energy for Fraunhofer diffraction pattern of 2D QD phase gratings with working phases of (a) 1.99999rad; (b) 2.00777rad.}
	\label{Fig:ContourQD}
\end{figure}
Since the rough 1D model gives a working phase of 2.00777rad \cite{Yan2013}, an initial interval of $[1.9, 2.1]$ is set in bisection algorithm. Then an optimized phase of 1.99999rad is obtained after 18 iterations. As Figure \ref{Fig:ContourQD} shown, being illuminated by a normally incident, unit-amplitude and monochromatic plane wave, the energy distributions across the QD grating at both phases look similar as anticipated. However, the energy difference between zeroth and first orders are $6\times 10^{-6}$ and $1.7\%$ at working phases of 1.99999rad and 2.00777rad, respectively. When applying the 1D model for the phase design of a QD grating with non-moderate parameters, especially if a big value of $W_{20}$ is selected (say $100\lambda$), the energy imbalance between first three orders will be even worse.  

\begin{figure}[t]
	\centering
  \includegraphics[width=0.4\textwidth]{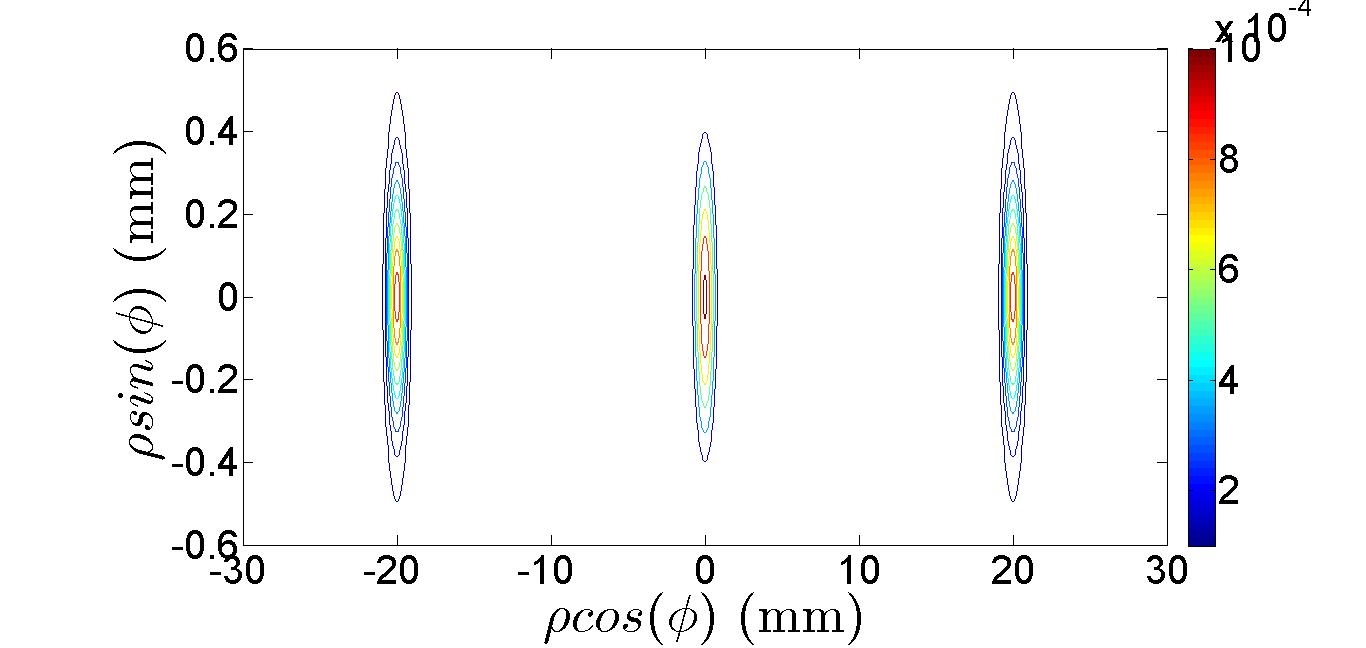}
	\caption{Contour plot of normalized energy for Fraunhofer diffraction pattern of the quasi-straight-line QD grating.}
	\label{Fig:ContourStrQD}
\end{figure}

Given that the Fraunhofer diffraction at a circular sector and the 2D mathematical model of QD grating are first developed in this paper, our theories and algorithms should be verified in practice. Here a quasi-straight-line QD grating, which reserves the parameters shown in Table \ref{Tbl:ParaQD} but sets $W_{20}$ to be $0.5\lambda$, is applied in the 2D QD grating model and a working phase of 2.00831rad is obtained. The Fraunhofer diffraction pattern is demonstrated in Figure \ref{Fig:ContourStrQD}, in which the positions of the peaks are identical with those derived by classic theory of diffraction grating \cite{Born2001} and the energy distribution tends to be the same with that of straight-line and period-fixed grating.

In conclusion, we have established an elaborate 2D analytic model of QD grating and obtained the Fraunhofer diffraction pattern. This model can be extended to the design of crossed QD grating for simultaneous 9-plane imaging. Beyond the design of grating, it can also be utilized in the design and optimization of simultaneous multi-plane imaging system. An updated model involved with a chromatic correction scheme using grisms \cite{Yan2013OE} is in progress and high order aberrations, i.e. spherical aberration and coma, will be considered in the near future.\\

\noindent \textbf{\Large{Funding.}} This work at HWU was funded by the Science and Technology Facilities Council (STFC). YF was funded by SUPA Prize Studentship and research grant of Prof. Xiaohong Fang of ICCAS. YF is currently funded by the France BioImaging (FBI) infrastructure ANR-10-INBS-04 (by Dr. Nadine Peyri\'{e}ras).

%\bigskip

% Bibliography
%\bibliography{sample}

\newpage
\renewcommand \refname{Full References}

% Please include bios and photos of all authors for aop articles 
\ifthenelse{\equal{\journalref}{aop}}{%
\section*{Author Biographies}
\begingroup
\setlength\intextsep{0pt}
\begin{minipage}[t][6.3cm][t]{1.0\textwidth} % Adjust height [6.3cm] as required for separation of bio photos.
  \begin{wrapfigure}{L}{0.25\textwidth}
    \includegraphics[width=0.25\textwidth]{john_smith.eps}
  \end{wrapfigure}
  \noindent
  {\bfseries John Smith} received his BSc (Mathematics) in 2000 from The University of Maryland. His research interests include lasers and optics.
\end{minipage}
\begin{minipage}{1.0\textwidth}
  \begin{wrapfigure}{L}{0.25\textwidth}
    \includegraphics[width=0.25\textwidth]{alice_smith.eps}
  \end{wrapfigure}
  \noindent
  {\bfseries Alice Smith} also received her BSc (Mathematics) in 2000 from The University of Maryland. Her research interests also include lasers and optics.
\end{minipage}
\endgroup
}{}

\end{document}